\documentstyle[mncite,draft,psfig]{mn}

\title[K-band Spectroscopy of XY Ari]{K-band spectroscopy of the intermediate
polar XY~Ari}

\author[S.\,P.\ Littlefair et.\,al.]{S.\,P.\ Littlefair,$^{1}$
 V.\,S.\ Dhillon,$^{1}$ T.\,R.\ Marsh$^2$ \\
$^1$Department of Physics and Astronomy, University of Sheffield, 
Sheffield S3 7RH, UK \\
$^2$Department of Physics and Astronomy, University of Southampton, 
Highfield, Southampton SO17 1BJ, UK \\}

\date{\center{\Large Submitted for publication in the Monthly Notices of the
Royal Astronomical Society \\ 
\vspace{.5cm} \today}} 

\begin{document}
\maketitle

\begin{abstract} 
We present the K-band infrared spectrum of the intermediate polar XY
Ari. The spectrum confirms the cataclysmic binary nature of XY Ari,
showing emission lines of He\,{\small I} ($\lambda$2.0587 $\mu$m) and
the Brackett and Paschen series of H\,{\small I}. The broad nature of
these lines suggest an origin in an accretion disc. The spectrum is
strongly reddened by absorption within the molecular cloud Lynds 1457
and shows prominent absorption features from the secondary star, from
which we determine a spectral type for the secondary of M0V. The
secondary contributes $80\pm6$\% of the K-band light. We derive a
visual extinction to XY~Ari of $A_v=11.5 \pm0.3$ and a distance of
$d=270 \pm 100$ pc, placing XY~Ari behind the molecular  cloud.
\end{abstract} 

\begin{keywords} 
binaries: close -- stars: individual: XY~Ari --  intermediate polars,
accretion, cataclysmic variables -- infrared: stars
\end{keywords}

\section{Introduction}
\label{sec:introduction}
XY~Ari (formerly known as H0253+193) was discovered as an X-ray source
using the {\em Einstein} satellite \cite{halpern87}. It lies within
$5^{\prime}$ of the core of the molecular cloud Lynds 1457 and was
initially suggested to be an X-ray emitting protostar. However, the
discovery of X-ray pulsations with a 206-s period \cite{takano89}
suggested that the X-rays originated from accretion onto a strongly
magnetised, rotating compact object. \scite{patterson90} noted that
the X-ray properties of XY~Ari strongly resemble those of an
intermediate polar (IP) and the binary nature of the system was
confirmed with the discovery of X-ray eclipses recurring with a 6.06-h
orbital period \cite{kamata91}. XY~Ari is the only intermediate polar
to exhibit deep X-ray eclipses.

High extinction through the molecular cloud means that XY~Ari is not
observable at optical wavelengths. The infrared (IR) counterpart was
identified by \scite{zuckerman92}. Infrared photometry was undertaken
by \scite{allan96} who modelled the ellipsoidal variations and found a
mass ratio in the range \mbox{ $0.43 < q < 0.71$} and an inclination in the
range \mbox{$80^{\circ} < i < 87^{\circ}$}. Here we present the IR 
(K-band) spectrum of XY~Ari. We use the spectrum to determine the spectral
type and contribution of the secondary star, as well as the distance and  
extinction to XY~Ari.

\section{Observations}
\label{sec:observations} 
The spectrum of XY~Ari presented in this paper was obtained on the
night of 1995 October  20 with the Cooled Grating Spectrometer 4
(CGS4) on the 3.8~m United Kingdom Infrared  Telescope (UKIRT) on
Mauna Kea, Hawaii. CGS4 is a 1--5  micron spectrometer containing an
InSb array with 256$\times$256 pixels. The 75~l\,mm$^{-1}$ grating in
the first order with the 150~mm camera gave a resolution of
approximately 350~km\,s$^{-1}$ and a wavelength range of approximately
0.7$\mu$m.

Optimum spectral sampling and bad pixel removal were obtained by
mechanically  shifting the array over two pixels in the dispersion
direction in  steps of 0.5 pixels. We employed the non-destructive
readout mode of the detector to reduce the readout noise. In order to
compensate for fluctuating atmospheric  OH$^-$ emission lines we took
relatively short exposures (typically 30 seconds) and nodded the
telescope primary so that the object spectrum  switched between two
different spatial positions on the detector.  The slit width was 1.2
arcseconds (projecting to approximately 1 pixel on the detector) and
was oriented at the parallactic angle. The seeing disc was equal to or
slightly larger than the slit width throughout. The observations were
made in photometric conditions.

A full journal of observations is presented  in table~\ref{tab:journal}. 
\begin{table}
\caption[]{Journal of observations. The spectral types of the dwarf stars
have been taken from the catalogue of \scite{kirkpatrick91} unless otherwise 
noted.}
\begin{tabular}{@{\extracolsep{-2.15mm}}lccccc}
& & & & \\
\hline
& & & & \\
Object &
\multicolumn{1}{c}{Spectral} &
\multicolumn{1}{c}{Date} &
\multicolumn{1}{c}{UTC} &
\multicolumn{1}{c}{UTC} &
\multicolumn{1}{c}{Exposure time} \\
 & \multicolumn{1}{c}{Type} & &
\multicolumn{1}{c}{start} &
\multicolumn{1}{c}{end} &
\multicolumn{1}{c}{(s)} \\ 
& & & & \\
XY Ari        &         & 20/10/95 & 09:07 & 12:22 & 5700 \\ 
Gl 775        & K5V     & 20/10/95 & 06:20 & 06:27 &   96 \\ 
Gl 764.1b     & K7V     & 20/10/95 & 05:43 & 05:52 &  240 \\ 
Gl 154        & M0V$^a$ & 21/10/95 & 09:39 & 09:48 &  192 \\ 
Gl 229        & M1V     & 06/02/96 & 05:00 & 05:08 &   72 \\ 
& & & & \\
\hline
& & & & \\
\multicolumn{6}{l}{$^a$\scite{hawley96}} \\
\end{tabular}
\label{tab:journal}
\end{table}

\section{Data Reduction} 
\label{sec:datared} 
The initial steps in the reduction of the 2D frames were performed
automatically by the CGS4 data reduction system \cite{daley94}.  These
were: the application of the bad pixel mask, bias and dark frame
subtraction, flat field division, interlacing integrations taken at
different detector positions, and co-adding and subtracting nodded
frames. Further details of the above procedures may be found in the
review by \scite{joyce92}. In order to obtain 1D data, we removed the
residual sky by subtracting a polynomial fit and then extracted the
spectra using an optimal extraction technique  \cite{horne86a}.  The
next step was the removal of the ripple arising from variations in the
star brightness between integrations (i.e. at different detector
positions). These variations were due to changes in the seeing, sky
transparency and the slight motion of the stellar image relative to
the slit.

There were two stages to the calibration of the spectra. The first was
the calibration of the wavelength scale using krypton arc-lamp
exposures. We had no readable arc frames from the night of 1995 October 20,
so an arc frame from the next night was used. A second-order
polynomial fit to the arc lines yielded an error of less than 0.001
microns (rms) and the error in the fit showed no systematic trend with
wavelength. The final step in the spectral calibration was the removal
of telluric  features and flux calibration. This was performed by
dividing the spectra to  be calibrated by the spectrum of an F-type
standard star, with its prominent  stellar features interpolated
across. We then multiplied the result by the  known flux of the
standard at each wavelength, determined using a black body  function
set to the same effective temperature and flux as the standard. As
well as correcting for the spectral response of the detector, this
procedure also  removed telluric absorption features from the object
spectra.

\section{Results}
\label{sec:results}
\begin{figure*}
\centerline{\psfig{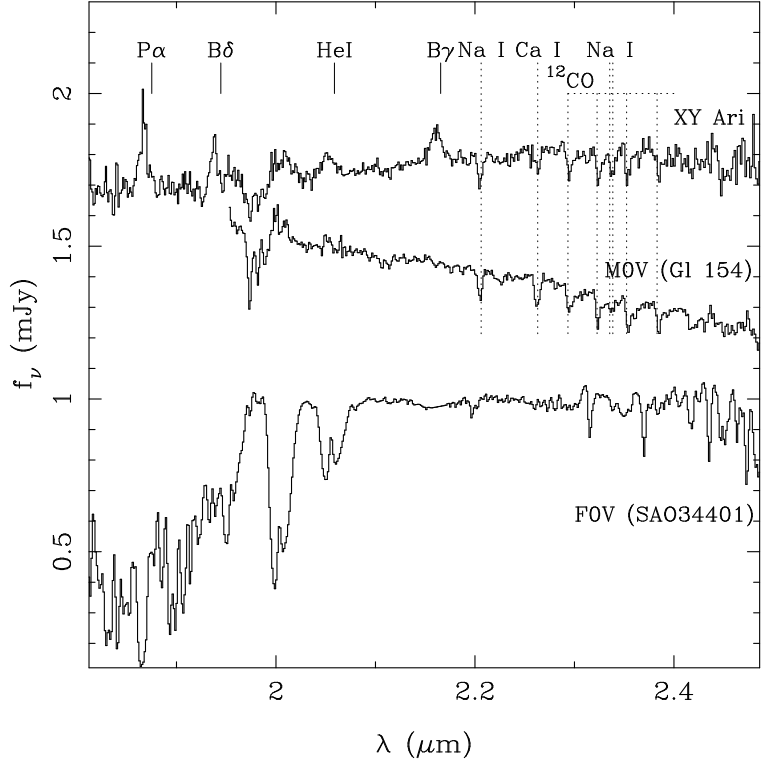}}
\caption{K-band spectra of the IP XY~Ari, and the M-dwarf Gl154. The spectra 
have been normalized by dividing by the flux at 2.24 $\mu$m and then offset 
by adding a multiple of 0.4 to each spectrum. Also shown is the spectrum of 
an F0V star, normalized by dividing by a spline fit to the continuum, 
which indicates the location of telluric features.}
\label{fig:spectra}
\end{figure*}
Figure~\ref{fig:spectra} shows the K-band spectra of the IP XY~Ari and
the  M0-dwarf Gl~154. The spectrum of XY~Ari shows strong, broad
emission lines of He\,{\small I} ($\lambda$2.0587 $\mu$m) and the
Brackett and Paschen series of H\,{\small I}, confirming the
cataclysmic variable nature of this system. The large velocity widths
of these lines imply an accretion disc origin, consistent with the
study of  \scite{allan96}, who found evidence for IR-emitting gas
around the white dwarf from the H-band lightcurve.  The spectrum also
shows strong absorption features of Na\,{\small I}, Ca\,{\small I} and
$^{12}$CO due to the secondary star   Note that the atomic features in
reality consist of a blend from many atomic species \cite{wallace96}
and the identification given here is that of the strongest contributor.

\subsection{Secondary star spectral type}
In order to determine the spectral type of the secondary star in
XY~Ari, and to determine its contribution to the K-band flux, we used
an optimal subtraction technique (e.\,g.\ \pcite{dhillon93}). Before
this could be carried out however, we needed to account for reddening
by the molecular  cloud, as this could affect the apparent spectral
type of the secondary star by altering the relative strengths of the
secondary star absorption features. To account for this we de-reddened
the spectra of XY~Ari using the  ``standard'' extinction curve of
\scite{howarth83} and colour excesses of $2<E(B-V)<5$. The
optimal subtraction technique described below was applied to the
dereddened spectrum in order to find the spectral type of the secondary
and its contribution to the K-band flux. It was found that, within the
range of colour excess used, the dereddening did not have a
significant effect on the results obtained. Therefore, in order to
determine the spectral type a colour excess of $E(B-V)=4$ was
used. This value of the colour excess produced a dereddened spectrum
whose continuum slope closely matched that of the late-K and early-M
dwarfs.

The optimal subtraction technique works as follows; first we normalised
the spectra of XY~Ari and the spectral type template stars by dividing
by a first-order polynomial fit to the continuum. A constant times the
normalised template spectrum was then subtracted from the normalised
spectrum of XY~Ari and the constant adjusted so as to minimise the
residual scatter in regions containing secondary star features. The
scatter was calculated by carrying out the subtraction and then
computing the $\chi^2$ statistic between the residual spectrum and a
smoothed version of this residual. Prior to the subtraction, the
template spectra should ideally be broadened to account for the rotational
velocity of the secondary star; the low velocity resolution of our
data made this step uneccessary. The value obtained for the percentage
contribution naturally depends on spectral type, the correct spectral
type being the one which minimises the value of $\chi^2$. The error in
the percentage contribution is the formal error calculated from the 
variation in $\chi^2$.

\begin{figure*}
\centerline{\psfig{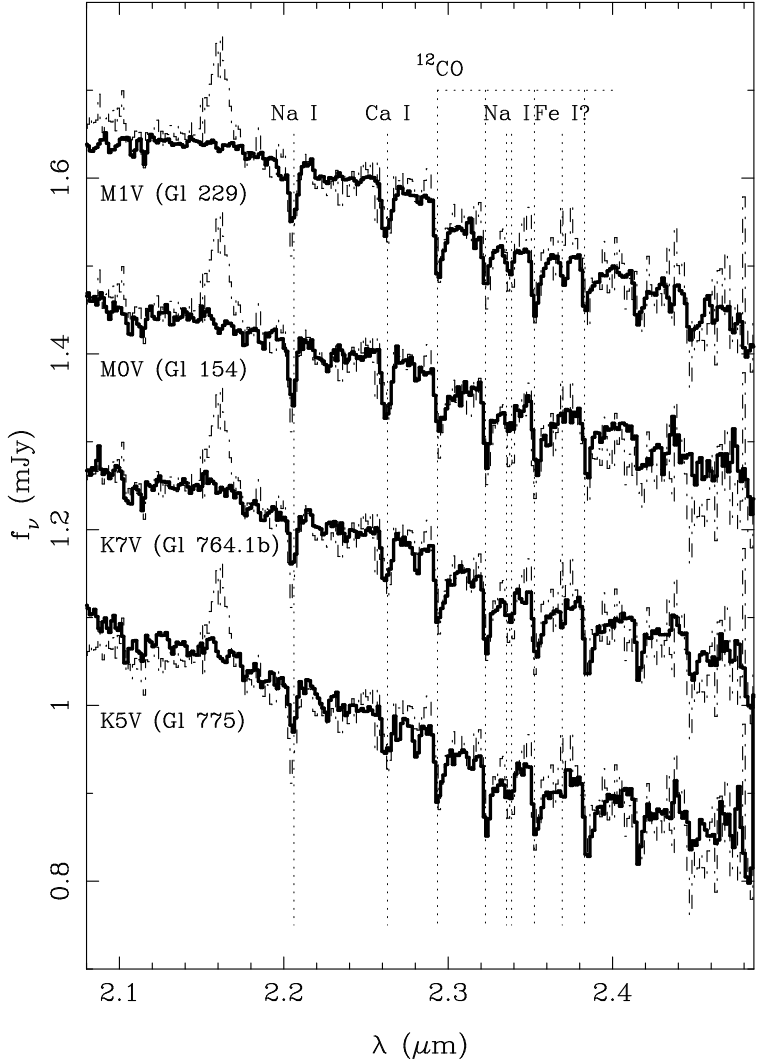}}
\caption{Fits to the K-band spectra of the IP XY~Ari. Template spectra
with a flat continuum added according to the results of the optimal
subtraction are shown in bold, over-plotted on the spectrum of XY~Ari,
plotted with a dashed line.}
\label{fig:fits}
\end{figure*}

For our data, however, the value of $\chi^2$ was not a revealing
statistic, as several spectral type templates produced
reduced-$\chi^2$ values of less than one. In order to select the
best-fitting spectral type the results of the optimal extraction
technique were examined by-eye. In order to  do this, we added a flat
continuum to the spectral-type template spectrum, so that the
template contributed the same amount to the K-band flux as was
suggested by the optimal subtraction technique. The resulting spectra
are shown in figure~\ref{fig:fits}, over-plotted on the spectrum of
XY~Ari. From this figure, we can see that the continuum slope around 2.1$\mu$m
is well fitted by the K7V and M0V stars, but poorly
fitted by the K5V and M1V stars. In selecting between the K7V and M0V
fits we have given more weight to the features bluewards of
$\sim$2.38$\mu$m as the features redwards of this wavelength are
strongly affected by telluric absorption. Hence we determine a that
the secondary star in XY~Ari has a spectral type of {\small
M}0$\pm$0.5{\small V} by inspection from figure~\ref{fig:fits}.
For a spectral type of M0{\small V} the optimal subtraction technique
outlined above indicates that the secondary star contributes $80\pm6$\% of 
the K-band flux. The majority of the remainder of the K-band flux
probably originates in the accretion disc.

\subsection{The extinction}
\label{subsec:ext}
If we denote the intrinsic apparent magnitude (i.\,e.\ the apparent
magnitude we would observe if there were no extinction) as
$m_{\lambda}^i$ and the absolute magnitude as $M_{\lambda}$ then
\begin{equation}
m_{\lambda}^i - M_{\lambda} = 5 \log_{10} \left( \frac{d}{10} \right)
\label{eq:1}
\end{equation}
where $d$ is the distance in pc. If the observed apparent magnitude is
$m_{\lambda}^o$, then
\begin{equation}
m_{\lambda}^o = m_{\lambda}^i + A_{\lambda}
\label{eq:2}
\end{equation}
where $A_{\lambda}$ is the extinction at wavelength $\lambda$. 
Equations~(\ref{eq:1}) \& (\ref{eq:2}) imply
\begin{equation}
m_K^o - m_J^o = (A_K - A_J) + (M_K - M_J)
\label{eq:3}
\end{equation}
which, using the extinction curve of \scite{howarth83} can be re-written
\begin{equation}
m_K^o - m_J^o = (M_K - M_J) -0.5 E_{(B-V)}
\label{eq:4}
\end{equation}

We can apply this to the secondary star in XY~Ari in order to
determine the extinction.  For a secondary of type early-M, the
secondary star in a CV contributes  roughly the same percentage of the
J- and K-band light  (see, for example, figure~5 in
\pcite{dhillon97a}). We therefore assume that the secondary
contributes 80\,\% of the J- and K-band flux in XY~Ari. Combined with
the J- and K-band magnitudes for XY~Ari from \scite{zuckerman92} of
$J=16.0$ and $K=13.3$ this gives  $m_K^o = 13.6$ and $m_J^o = 16.3$
for the observed apparent magnitudes of the secondary star in
XY~Ari. From \scite{bessel91} the absolute K- and  J- band magnitudes
for a {\small M}0{\small V} star are $M_K = 5.22$ and  $M_J =
6.06$. Assuming the error in the extinction is dominated by the error
in estimating the spectral type of the secondary star in XY~Ari, this
gives a colour excess of $E(B-V)=3.7\pm 0.1$ or a visual extinction of
$A_V=11.5\pm 0.3$.

\subsection{The distance to XY~Ari}
\label{subsec:dist}
The distances to CVs can be measured from K-band spectra using a
modification of a method first proposed by \scite{bailey81}. The
distance modulus can be written in terms of the K-band surface
brightness of the secondary star, $S_k$, as
\begin{equation}
S_k = m_K^o - A_K + 5 - 5\log d + 5 \log (R/R_{\odot})
\label{eq:5}
\end{equation}
where $A_K$ is the K-band extinction and $R$ is the
radius of the secondary star. $m_K^o$ was determined in 
section~\ref{subsec:ext}. Using the extinction derived in 
section~\ref{subsec:ext} and the extinction curve of \scite{howarth83} we
estimate $A_K = 1.3 \pm 0.1$. Assuming the error in $V-K$ to be dominated by
the error in estimating the spectral type of the secondary star in XY~Ari,
\scite{bessel91} gives a $V-K$ colour of $V-K = 3.7\pm0.4$. Using the
calibrations of \scite{ramseyer94} this gives $S_k = 4.4 \pm0.1$. We have
estimated the radius of the secondary star to be $R/R_{\odot} = 0.7 \pm0.2$
from the orbital period-radius relation (equation 11) given by \scite{sad98}.
These values, in conjunction with equation~\ref{eq:5} gives a distance to
XY~Ari of $270 \pm 100$ pc. This places XY~Ari well behind the molecular 
cloud, for which \scite{hobbs88} derived a distance of $d < 110 pc$.

By assuming the light from XY~Ari consisted entirely of light
from a secondary star of spectral type K8, \scite{zuckerman92} 
found their J- and K-band colours were consistent with a distance of
200 pc and a visual extinction of $A_V = 11.5$ mag. Our values
are in agreement with these results.

\section{Conclusions}
\label{sec:conclusions}
The K-band spectrum of XY~Ari shows strong, broad emission lines, confirming
the cataclysmic variable nature of the object. The spectrum is strongly 
reddened and shows prominent absorption features from the secondary star.
We estimate the spectral type of the secondary star to be M0V. The secondary
star contributes $80\pm6$\% of the K-band flux. Comparing the observed
infrared colours with the intrinsic colour of the secondary star we estimate
a distance to XY~Ari of $270 \pm 100$ pc and a visual extinction of 
$A_V = 11.5 \pm 0.3$, placing XY~Ari behind the molecular cloud Lynds 1457.

\section*{\sc Acknowledgements}
SPL is supported by a PPARC studentship. 
UKIRT is operated by the Joint Astronomy Centre on behalf of the Particle 
Physics and Astronomy Research Council. The authors acknowledge the data 
analysis facilities at Sheffield provided by the Starlink Project which is run
by CCLRC on behalf of PPARC. 
\bibliographystyle{mnras}
\bibliography{refs}

\end{document}